\documentclass{article}
\usepackage{spconf,amsmath,graphicx}
\usepackage{bm}
\usepackage{amssymb}
\usepackage{url}
\usepackage{graphicx, multirow, booktabs} % Required for tables
\usepackage{enumitem}
\setlist{nosep, leftmargin=14pt}
\usepackage{placeins}
\usepackage{mwe} % to get dummy images

\title{Informed Bootstrap Augmentation Improves EEG Decoding}
\makeatletter
\def\@name{\emph{Woojae Jeong$^{1,2}$, Wenhui Cui$^{1}$, Kleanthis Avramidis$^{3}$,} 
  \\ \emph{Takfarinas Medani$^{1}$, Shrikanth Narayanan$^{1,3}$, and Richard M. Leahy$^{1}$}}
\makeatother

\address{\\ $^{1}$ Ming Hsieh Department of Electrical and Computer Engineering, University of Southern California \\ $^{2}$ Alfred E. Mann Department of Biomedical Engineering, University of Southern California \\ $^{3}$ Thomas Lord Department of Computer Science, University of Southern California}

\begin{document}
%\ninept
%
\maketitle
\begin{abstract}
Electroencephalography (EEG) offers detailed access to neural dynamics but remains constrained by noise and trial-by-trial variability, limiting decoding performance in data-restricted or complex paradigms. Data augmentation is often employed to enhance feature representations, yet conventional uniform averaging overlooks differences in trial informativeness and can degrade representational quality. We introduce a weighted bootstrapping approach that prioritizes more reliable trials to generate higher-quality augmented samples. In a Sentence Evaluation paradigm, weights were computed from relative ERP differences and applied during probabilistic sampling and averaging. Across conditions, weighted bootstrapping improved decoding accuracy relative to unweighted (from $68.35\%$ to $71.25\%$ at best), demonstrating that emphasizing reliable trials strengthens representational quality. The results demonstrate that reliability-based augmentation yields more robust and discriminative EEG representations. The code is publicly available at \url{https://github.com/lyricists/NeuroBootstrap}
\end{abstract}
\section{Introduction}
\label{sec:intro}

Electroencephalography (EEG) provides a powerful, noninvasive method for examining the temporal dynamics of neural processing and has been widely used to decode cognitive and perceptual states from brain activity~\cite{baillet_electromagnetic_2001, cichy_multivariate_2017}. However, EEG signals are inherently noisy and highly variable across trials and participants, often requiring repeated stimulus presentations and trial averaging to obtain stable neural representations~\cite{luck2005ten}. But such repetition is not always feasible in complex cognitive paradigms or naturalistic experiments.

These constraints pose challenges for machine learning and deep learning approaches, which depend on sufficient data volume, controlled variability, and stable feature representations to learn discriminative patterns effectively. Decoding often degrades under high variability and limited data, conditions that are common in complex EEG tasks~\cite{saeidi2021neural, xu2020cross, bomatter2025limited, roy2019deep}. Augmentation strategies based on bootstrapping with sub-averaging have emerged as practical solutions for obtaining more stable representations from limited data~\cite{sieluzycki2021reducing}. However, when some trials carry stronger task-relevant information than others, uniform averaging can dilute informative patterns by weighting all samples equally, resulting in a weaker representation.

To address this limitation, we propose a weighted bootstrapping augmentation framework that generates reliable representations of EEG signals by probabilistically sampling and averaging trials based on their estimated reliability. Unlike conventional uniform bootstrapping, our method emphasizes more informative trials, derived from task-specific measures, allowing us to create more robust averaged samples that preserve meaningful task-related neural characteristics.

We evaluate the effectiveness of this weighted augmentation in a Sentence Evaluation paradigm~\cite{hughes2025precog}. Weighted bootstrapping consistently yielded higher decoding accuracy than both uniform and random-weighted approaches, indicating that emphasizing trial-level reliability sharpens the resulting feature representations and more effectively preserves informative neural structure despite noise or variable signal quality. Our findings highlight the potential of weighted bootstrapping as a simple tool to extend EEG decoding capabilities.

\section{Materials and Methods }
\label{sec:method}

\subsection{PRECOG Dataset}
\label{ssec:dataset}
\textbf{Dataset}: EEG and behavioral data from a total of 137 subjects (college students, aged 18\textendash 25 years) with varying mental health profiles obtained from the \textit{Multimodal Integration of Neural and Biobehavioral Signals for Predicting Preconscious Responses} (PRECOG) study~\cite{hughes2025precog} were used in this work. The PRECOG study investigated the neurophysiological mechanisms underlying depression and suicidal ideation through a series of cognitive tasks (Emotional Stroop and Sentence Evaluation task). In the present study, data from the Sentence Evaluation task were analyzed. Comprehensive details of the experimental protocols are provided in~\cite{hughes2025precog}.
% Subjects were screened into three groups based on the Patient Health Questionnaire-9 (PHQ-9)~\cite{kroenke2001phq} and the Suicidal Ideation Scale (SIS)~\cite{rudd1989prevalence} scores: Group 1 - healthy control (C) \(N=48)\), Group 2 - depressed and non-suicidal (D \(N=41\)), and Group 3 - depressed and suicidal (S \(N=48\)). 

\textbf{Sentence Evaluation task}: The task was designed to elicit cognitive responses to emotionally salient and self-referential statements~\cite{hughes2025precog}. During the Sentence Evaluation task, each subject viewed 160 self-referential sentences of positive, neutral, or negative sentiment, consisting of 80 base sentences, each paired with an alternative version that differed only in the final word. Each sentence contained 4 to 14 words, presented twice, resulting in a total of 320 trials per subject. Sentences were presented word by word on a black background, with each word displayed for 300 ms, with a 300 ms inter-word interval (IWI). The final word was presented for 600 ms, followed by a 300-ms IWI and a 2-s response window. Subjects were instructed to indicate whether they agreed or disagreed with each statement. The task consisted of 4 blocks, each containing 80 trials. The full set of 160 sentences was randomly presented in blocks 1 and 2 and repeated in blocks 3 and  4 with independent randomization.

\textbf{Sentence Category}: Sentences were categorized into four topics of interest (TOIs). Neutral Biographical, Depressive Reflection, Depressive Action, and Suicidal Intention. This study focuses on the Neutral Biographical and Suicidal Intention TOIs, which were designed to serve as the baseline and to elicit the strongest emotional responses, respectively. Each sentence was labeled as Type 1 or Type 2 based on the expected subject response, with the two types intentionally constructed to elicit distinct semantic and emotional neural activity. The Neutral Biographical category included sentences reflecting everyday experiences typical of college students (e.g., Type 1: \textit{“I have weekly get together with students”}; Type 2: \textit{“I have weekly get together with lawyers”}), whereas the Suicidal Intention category comprised sentences associated with suicidal thoughts or desires (e.g., Type 1: \textit{“The thought of ending my life is relieving”}; Type 2: \textit{“The thought of ending my life is ridiculous”}). For clarity, we refer to the biographical and intention topics as $Bio$ and $Int$, respectively, and $BI$ denotes the combined dataset. 

\textbf{EEG Preprocessing}: EEG data were recorded using a 64-channel active electrode system (actiCAP, Brain Products, GmbH) with a 1 kHz sampling rate. Preprocessing was performed in Brainstorm software~\cite{tadel_brainstorm_2011} and EEGLab~\cite{delorme_eeglab_2004}. We removed the powerline noise (60 Hz) using a notch filter and then applied a bandpass between 0.5\textendash 80 Hz. The artifact Subspace Reconstruction (ASR) routine~\cite{mullen_real-time_2015} was used to remove noisy channels, which were subsequently interpolated by averaging the neighboring channels within a 5 cm radius. Independent Component Analysis (ICA) was used to remove ocular, heart, and muscle artifacts. EEG signals were re-referenced to the common average, detrended, and epoched from -200 ms to 1500 ms from the stimulus onset. Each trial was z-scored using the pre-stimulus baseline (-200\textendash 0 ms), then low-pass filtered at 20 Hz with a zero-phase IIR filter and downsampled to 250 Hz. In this study, we excluded 17 outer channels (Fp1, Fp2, AF7, AF8, F7, F8, FT7, FT8, FT9, FT10, T7, T8, TP7, TP8, TP9, TP10, and Iz) prone to noise.

\subsection{Feature Quality Evaluation metrics}
\label{ssec:quality}

\textbf{SNR}: Signal-to-noise ratio (SNR) was evaluated as~\cite{parks2016bootstrap}:
\begin{equation}
    SNR=20*log\frac{RMS(ERP_{signal})}{RMS(ERP_{baseline})},
\end{equation}
where $RMS(ERP_{signal})$ is the root mean square of ERP signal within the signal window, and $RMS(ERP_{baseline})$ represents the root mean square of baseline noise estimated from the pre-stimulus time window (-200-0 ms). The $ERP$ was estimated by averaging across all trials within each subject and then averaging across all subjects.

\textbf{ERP difference}: The ERP difference was computed as the mean amplitude difference between the Type 1 and Type 2 conditions:
\begin{equation}
\Delta ERP=\frac{1}{N_{t}}\Sigma_{t}(ERP_{Type1}(t)-ERP_{Type2}(t)), 
\end{equation}
where $N_{t}$ denotes time samples within the signal time window. $ERP_{Type1}$ and $ERP_{Type2}$ represent the averaged EEG across all trials for each sentence type. Within-subject average $\Delta ERP$ was then averaged across all subjects.

\subsection{Decoding with Bootstrap Augmentation}
\label{ssec:decoding}

\textbf{Weighted Bootstrapping}: Trial weights were computed from the $\Delta ERP$ ratio between the $Bio$ and $Int$ TOIs, with weight estimation performed using data from all other subjects so that, during within-subject decoding, the held-out subject's data never contributed to the weighting, as:
\begin{equation}
    w_{Bio}=1, w_{Int}=\frac{|\Delta ERP_{Int}|}{|\Delta ERP_{Bio}|}
\end{equation}
Each TOI contained $M$ trials, and weights were assigned uniformly within each TOI (i.e., all trials in the same TOI received the same weight) as:
\begin{equation}
\mathbf{W} = \bigl[\, w_{\mathrm{Bio}}\mathbf 1_M,\; w_{\mathrm{Int}}\mathbf 1_M \,\bigr]^{\top}
\end{equation}
During augmentation, $k$ trials were randomly selected with replacement according to these weights, with the normalized sampling probabilities as:
\begin{equation}
\mathbf{P} = \frac{1}{\sum_{j=1}^{2M} W_j } \mathbf{W},
\end{equation}
\begin{equation}
\mathbf{N}^{(b)} \sim \mathrm{Multinomial}\big(k;\mathbf{P}\big)
\end{equation}
The $b$-th augmented trial was obtained as the average of the selected trials:
\begin{equation}
\tilde{x}^{(b)} = \frac{1}{k}\sum_{i=1}^{2M} \mathbf{N}^{(b)} \, x_i
\end{equation}
This procedure was repeated $L$ times, resulting in $L$ augmented trials.

\begin{figure}[hbt]
\begin{minipage}[b]{1.0\linewidth}
  \centering
\centerline{\includegraphics[width=8.5cm]{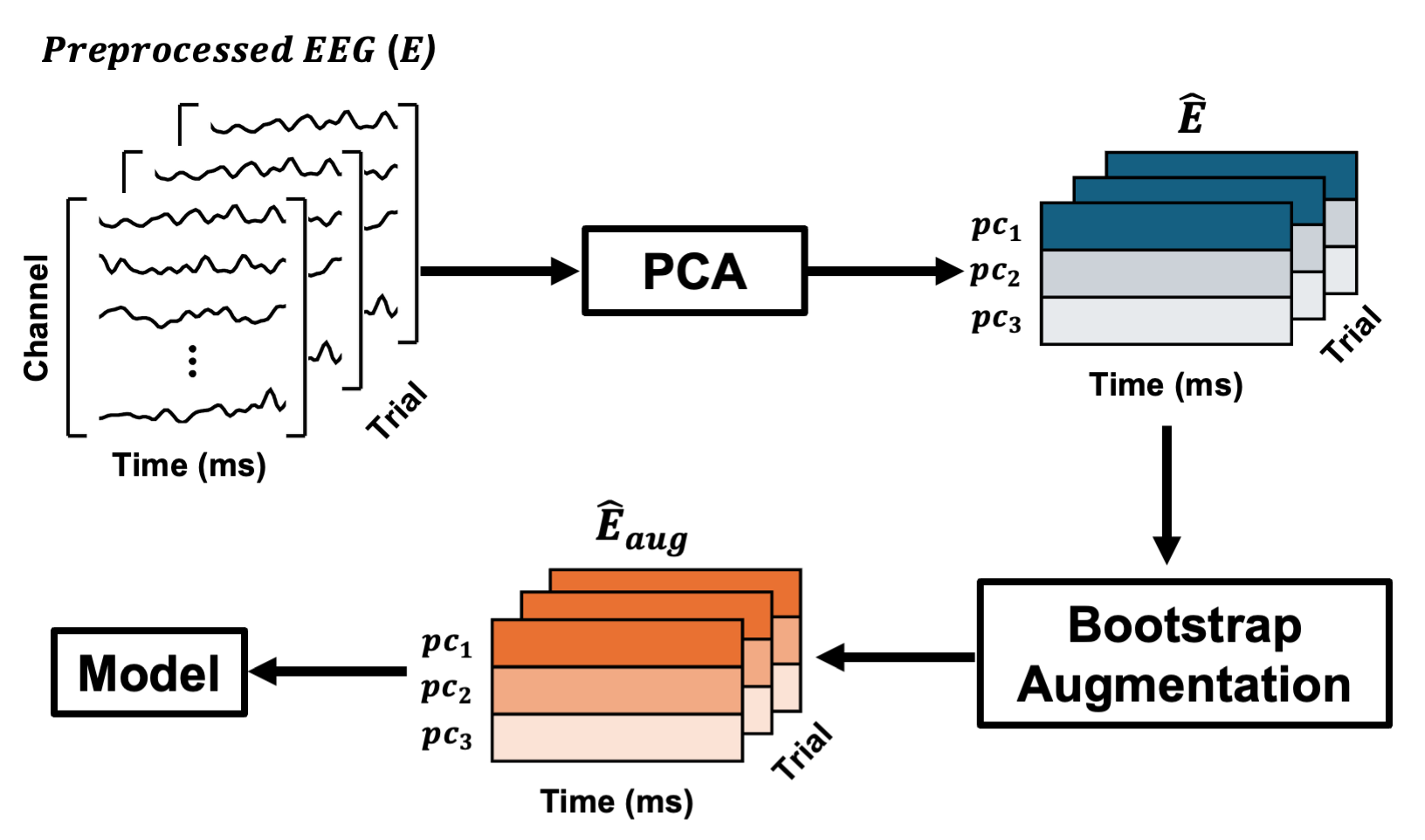}}
\end{minipage}
\caption{EEG within-subject sentence-type decoding pipeline. The first three PCs were extracted as spatial features ($\bm{\hat{\bm{E}}}$) through PCA across channels from the preprocessed EEG (\(E\)). Augmented trials ($\bm{\hat{\bm{E}}_{aug}}$) were then generated using the bootstrapping method.}
\label{fig:pipeline}
\end{figure}

\textbf{EEG Sentence-type Decoding Pipeline}: We followed the EEG decoding procedure (Fig.~\ref{fig:pipeline}) proposed in our previous study~\cite{jeong2025time}. Trials from each subject were divided into training and test sets using a stratified 5-fold cross-validation. To extract the shared spatial features across groups in a lower-dimensional space, we applied principal component analysis (PCA)~\cite{abdi2010principal} to the group-level average EEG responses concatenated over time across channels. Group-level averages were obtained by first averaging training trials within each subject and subsequently averaging these subject responses within each group. We computed the projection matrix $(\bm{W} \in \mathbb{R}^{3 \times 47})$ of the first three principal components (PCs). Each subject's EEG data was projected onto the latent feature space as:
\begin{equation}
    \bm{\hat{E}} = \bm{W E} \quad \text{where } \bm{\hat{E}} \in \mathbb{R}^{3 \times T \times L}
\end{equation}
where T is the time length and L is the number of augmented trials. 
For each subject, we generated 250 augmented trials (125 Type 1 and 125 Type 2) through bootstrap averaging within the respective training and test sets using two sampling schemes: (a) Uniform probability, (b) Weighted probability, and (c) random-shuffled probability as in Equation~4 \& 5. Finally, we performed binary sentence-type classification within each subject, using machine learning models.

\textbf{Model Training and Evaluation}: To evaluate our proposed augmentation scheme, we used a linear support vector machine (SVM)~\cite{hearst1998support} and DeepConvNet~\cite{hbm23730}. We performed within-subject sentence-type decoding using 5-fold cross-validation, and decoding accuracies were averaged across subjects. SVM used an 80\%-20\% (training/test) split, whereas DeepConvNet used an 80\%-10\%-10\% (training/validation/test) split for training and evaluation. DeepConvNet was trained using the Adam optimizer with a fixed learning rate of 0.001, batch size of 64, and 80 training epochs. Early stopping was applied with a patience parameter of 10 epochs, based on the validation loss, to prevent overfitting. The model that achieved the minimum validation loss was retained for evaluation.

\begin{figure}[hbt]
\begin{minipage}[b]{1.0\linewidth}
  \centering
\centerline{\includegraphics[width=8.5cm]{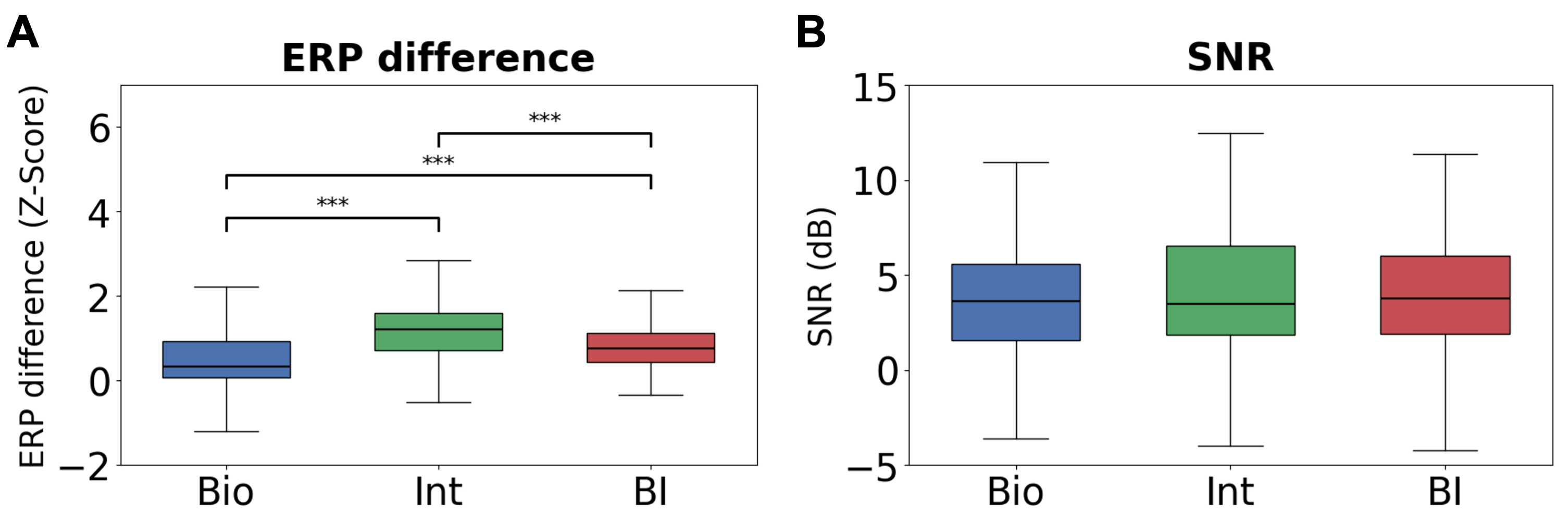}}
\end{minipage}
\caption{Feature quality evaluation. \textbf{A}. Mean ERP difference (Z-Score) across subjects for each condition. \textbf{B}. Mean SNR (dB) across subjects for each TOI. Stars indicate significant differences between conditions (paired t-test, $N=137$, *** $p<0.001$, FDR-corrected).}
\label{fig:SNR}
\end{figure}

% \vspace{-1.5ex}
\section{Experiments and Results}
\label{sec:results}

\subsection{Feature Quality Evaluation}
To quantitatively assess which TOIs were more effective, we evaluated ERP differences and SNR across TOIs using signals from the 300-600 ms post-stimulus window. The $Int$ TOI exhibited the largest ERP difference, significantly greater than both $BI$ and $Bio$ (Fig.~\ref{fig:SNR}A), while $BI$ was intermediate. No significant differences were observed in SNR across TOIs (Fig.~\ref{fig:SNR}B). These findings suggest that trials in the $Int$ TOI carry greater discriminative information for sentence-type decoding while the SNR across TOIs was kept the same.

\begin{figure}[hbt]
\begin{minipage}[b]{1.0\linewidth}
  \centering
\centerline{\includegraphics[width=8.5cm]{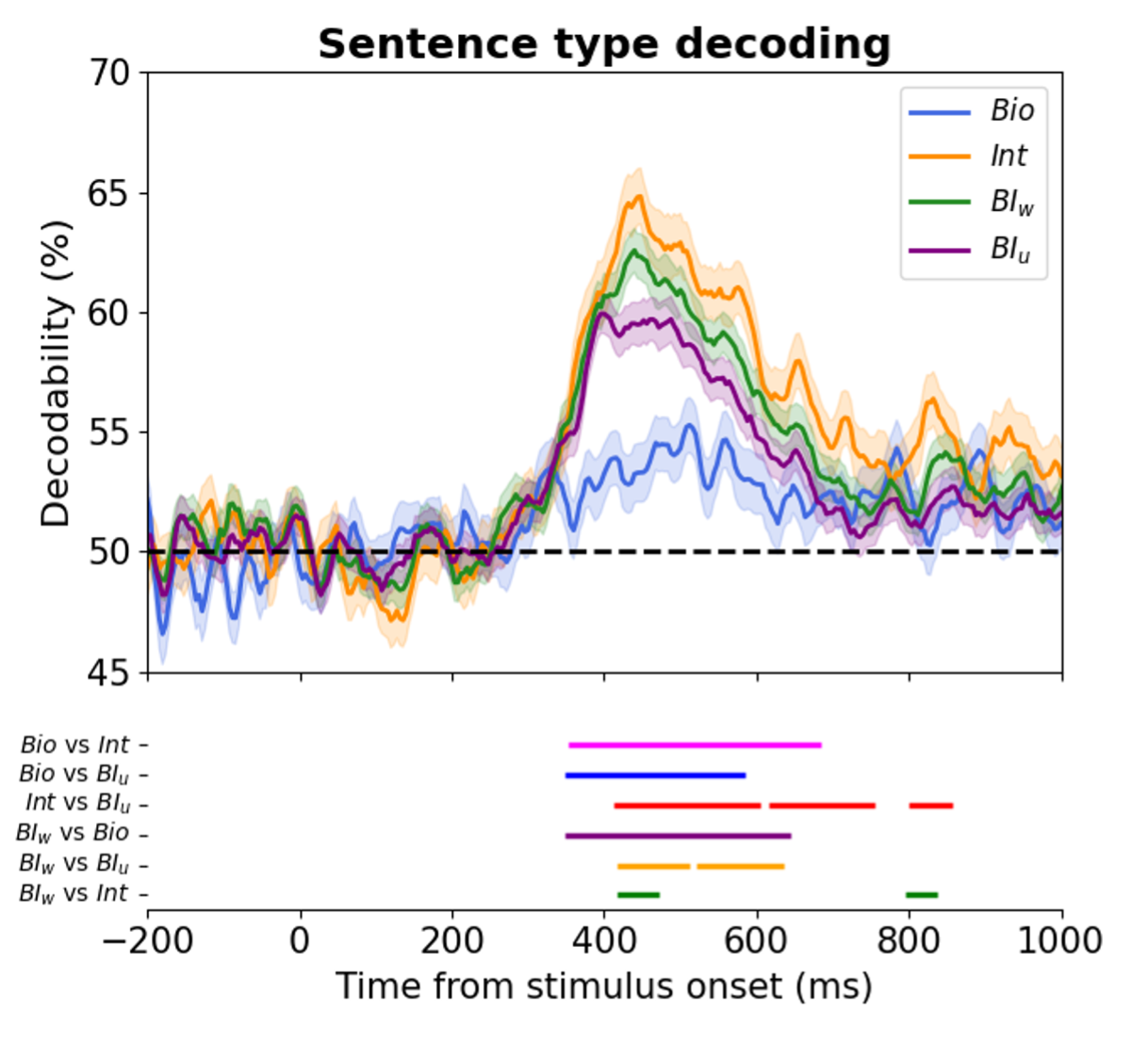}}
\end{minipage}
\caption{SVM based time-resolved within-subject EEG sentence-type decoding performance across conditions. Decoding accuracy is shown for $Bio$, $Int$, $BI_u$ with uniform bootstrapping, and for $BI_w$ with weighted bootstrapping. Shaded areas denote the standard error of the mean across subjects. Horizontal colored lines indicate time intervals with significant differences between paired conditions (using cluster-based permutation test, $N=137$, $p<0.05$).}
\label{fig:decoding}
\end{figure}

% \vspace{-1.5ex}
\subsection{Time-Resolved Sentence Type Decoding}
\label{ssec:subhead}

The number of source trials was matched across TOIs prior to augmentation. Both $Bio$ and $Int$ contained 80 trials each. The combined $BI$ set was formed by randomly selecting 40 trials from each TOI. Augmentation was performed by sub-averaging 8 trials to generate 250 samples per TOI condition, and sentence-type decoding was performed at each time $t$ using a linear SVM for each subject.

We observed robust sentence-type decoding across all conditions, with peak mean accuracy occurring between 300-600 ms after the stimulus onset (Fig.~\ref{fig:decoding}). Decoding performance followed the affective strength of the TOIs, highest for $Int$, lowest for $Bio$, and intermediate for $BI_u$ (uniform bootstrapping). Weighted bootstrapping ($BI_w$) significantly improved the decoding performance compared to the uniform bootstrapping ($BI_u$), indicating that emphasizing more informative trials during augmentation enhances decoding performance while preserving the temporal dynamics of sentence-related neural responses.

\subsection{Effect of Weighted Bootstrapping.}

To evaluate whether the weighted bootstrapping augmentation robustly improves decoding performance, we compared the sentence-type decoding accuracies obtained from uniform, weighted, and random-weighted bootstrapping using the DeepConvNet model. Based on the time-resolved decoding results, which showed a robust peak between 300\textendash 600 ms (Fig.~\ref{fig:decoding}), this time window was selected as the temporal feature input to the model. The mean decoding accuracy ($\pm SE$) averaged across subjects for each condition is reported in Table~\ref{tab:results}.

Overall, weighted bootstrapping significantly enhanced decoding performance (paired t-test, $N=137$, $p<0.05$, FDR-corrected) compared to both uniform bootstrapping and random weighted bootstrapping (Table~\ref{tab:results}). Random weighted bootstrapping did not differ significantly from uniform bootstrapping. Performance further improved when using a larger number of source trials (80$\rightarrow$160 trials) and greater sub-averaging sizes (8$\rightarrow$12$\rightarrow$16 trials). Under the 80-8 trial setting, decoding using only the $Int$ condition achieved the highest accuracy ($68.67\%\pm1.03$). However, the overall best performance was obtained with the 160-16 trial setting under weighted bootstrapping($71.25\%\pm1.03$).

Because the number of sub-averaging trials is constrained by the available source trials, the $Int$ condition alone was limited to a maximum sub-averaging size of 8. By combining $Int$ and $Bio$ trials to increase the pool of source samples, a larger sub-averaging size could be achieved, resulting in significantly higher decoding performance. These findings demonstrate that leveraging feature informativeness through weighted bootstrapping can result in consistent and measurable improvements in decoding accuracy.

% Single column
\begin{table}[t]
\centering
\footnotesize
\setlength{\tabcolsep}{4pt} % reduce column spacing
\renewcommand{\arraystretch}{1.2} % tighter rows
\resizebox{\columnwidth}{!}{ % scale to fit one column width
\begin{tabular}{cccccc}
\hline
 & \multicolumn{2}{c}{\textbf{Trial setting}} & \multicolumn{3}{c}{\textbf{Weights}} \\  % only span columns 2–4
\cline{2-6}
\textbf{TOI} & \textbf{Source} & \textbf{Sub-avg} & $w_u$ & $w$ & $w_r$ \\
\hline
\textbf{Bio}
 & $80$ & $8$ & $58.32\pm1.11$ & -- & -- \\

 \textbf{Int}
 & $80$ & $8$ & $68.67\pm1.05$ & -- & -- \\
\hline

\multirow{4}{*}{\textbf{BI}}
 & $80$ & $8$ & $62.76\pm0.95$ & $66.03\pm1.01$ & $63.49\pm0.89$ \\
 & $160$ & $8$ & $65.54\pm0.99$ & $67.78\pm1.01$ & $64.98\pm1.03$ \\
 & $160$ & $12$ & $68.08\pm1.02$ & $70.53\pm1.02$ & $67.70\pm1.00$ \\
 & $160$ & $16$ & $68.34\pm1.03$ & $\mathbf{71.25\pm1.03}$ & $67.32\pm1.03$ \\
\hline

\end{tabular}
}
\caption{DeepConvNet based mean sentence type decodability ($\pm$ SE, \%) averaged across subjects for each condition by bootstrapping type (\(w_u\): uniform bootstrapping, \(w\): weighted bootstrapping, and \(w_r\): random weighted bootstrapping) and trial setting (source trials and sub-averaging size). All conditions contained 250 augmented trials.}
\label{tab:results}
\end{table}

\section{Discussion}
\label{sec:discussion}
Weighted bootstrapping proved to be an effective augmentation strategy for EEG decoding, with data-driven reliability weights yielding consistently higher accuracy than uniform or random-weighted bootstrapping across all settings.

Previous work has shown that EEG decoding accuracy improves with averaging across more trials, which enhances signal stability and reduces noise contamination~\cite{lopez2020multivariate}. Our results are consistent with this. However, weighted bootstrapping yielded an additional performance gain beyond what can be attributed to signal averaging alone. This improvement arises from leveraging the informativeness of the trials---specifically, emphasizing trials that carry stronger task-relevant neural signatures---rather than merely increasing the SNR.

These findings highlight that weighted bootstrapping enhances augmentation efficiency by producing more reliable and informative synthetic samples, even under conditions of limited or noisy data. This capability is especially advantageous for complex experimental designs with constrained recording opportunities, where maintaining both data quality and quantity is difficult. By emphasizing more informative trials, the method strengthens feature representations and supports more robust and generalizable decoding.

Future work will extend this framework to subject-level classification and explore adaptive, model-driven weighting schemes to further optimize the selection of reliable samples.

\section{Compliance with ethical standards}
\label{sec:ethics}
All experimental protocols were approved by the Institutional Review Board (IRB) of USC (UP-23-00071).

\section{Acknowledgments}
\label{sec:acknowledgments}
This study was sponsored by the Defense Advanced Research Projects Agency (DARPA) under cooperative agreement No. N660012324006 and by the National Institute of Biomedical Imaging and Bioengineering (NIBIB) under award number R01EB026299. The content of the information does not necessarily reflect the position or the policy of the Government, and no official endorsement should be inferred. We thank Dr. Dimitrios Pantazis for his constructive feedback and insightful discussions on the study.

\bibliographystyle{IEEEbib}
\bibliography{refs}

\end{document}